\documentclass[%
reprint,
 amsmath,amssymb,
 aps,
floatfix,
]{revtex4-2}

\usepackage{hyperref}
\hypersetup{colorlinks,linkcolor={blue},citecolor={blue},urlcolor={red}} 
\usepackage{physics}

\usepackage{ulem}
\usepackage[dvipsnames]{xcolor}

\usepackage{subcaption}
\usepackage{graphicx}

\usepackage{float}
\usepackage[dvipsnames,table]{xcolor}  
\definecolor{mycolor}{rgb}{0.5,0.,0.5}

\usepackage{graphicx}
\usepackage{dcolumn}
\usepackage{bm}

\date{\today}
\begin{document}

\title{
 Optical smoothing broadens cross beam energy transfer resonance
}
 
\author{Y. Lalaire$^{1-2}$}\email{yann.lalaire@cea.fr}
\author{C. Ruyer$^{1-2}$} \email{charles.ruyer@cea.fr}
\author{A. Debayle$^{2-3}$} 
\author{G. Bouchard$^{1-2}$} 
\author{R. Capdessus$^{1-2}$} 
\author{A. Fusaro$^{1-2}$} 
\author{P. Loiseau$^{1-2}$} 
\author{L. Masse$^{1-2}$} 
\author{P. E. Masson-Laborde$^{1-2}$} 
\author{D. Bénisti$^{1-2}$}
\affiliation{$^1$CEA, DAM, DIF, F-91297 Arpajon, France}
\affiliation{$^2$Université Paris-Saclay, CEA, LMCE, 91680 Bruyère-Le-Chatel, France}
\affiliation{$^3$Focused Energy GmbH, Im Tiefen See 45, 64293 Darmstadt, Germany}

\begin{abstract}
We use the  theoretical framework introduced in the companion paper to provide simple formulas as regards the resonance conditions for CBET with smoothed laser beams. 
Our analytical CBET model with optical smoothing shows that these fusion-critical lasers produce a significantly broader resonance than conventional plane wave models predict.
%
%
In particular, temporal smoothing, as used in many high energy laser facilities, and flow components normal to the CBET ion acoustic waves, significantly modify the power transfer between smoothed beams.
Our model predicts that the energy transfer rate out of resonance is substantially higher with optical smoothing than without, a result that has profound implications for optimizing predicting and interpreting  future fusion experiments. We  provide a simple criterion which pinpoints the laser and plasma  parameters for which laser smoothing impacts CBET. These findings pave the way for experimental investigations in high-energy-density physics and fusion energy.
\end{abstract}

\maketitle

Significant fusion gains demonstrated by the National Ignition Facility (NIF) have driven important research initiatives toward inertial fusion energy (IFE) \cite{PRL_Fusion_2022}. The design of these fusion power plants requires highly efficient coupling between the laser drive and the fuel capsule, yet laser-plasma instabilities (LPI) remain a major challenge. LPI scatter pump energy in undesirable directions, reducing the energy coupling efficiency. Among the various LPI, Cross-Beam Energy Transfer (CBET) frequently occurs, impacting capsule symmetry and plasma properties, either in directly \cite{Craxton_2015} or indirectly driven experiments \cite{Kritcher2022,Hao_2025}. The prediction and understanding of CBET have been extensively studied, both theoretically \cite{Michel2009, Marion2016, Debayle2018, Huller2020, Seaton2022-1} and experimentally \cite{Neuville_2016, Neuville2018, Turnbull_2020, PRL_Dewald_2013, Debayle_2025}, with important implications for the design and interpretation of ICF experiments \cite{Colaitis_2016, Strozzi2017, Debayle2019, Follett2018}.
 
CBET occurs when the crossing of two laser beams with nearly identical wavelengths generates a grating that drives ion acoustic waves (IAWs) in the plasma through the ponderomotive force, driving power exchange between the laser beams. In both directly and indirectly driven capsules, CBET affects the symmetry of implosion and plasma properties \cite{Craxton_2015, Kritcher2022}. The power transfer is resonant when the phase velocity of the ponderomotive grating matches the sound speed in the plasma rest frame. This condition is met in various scenarios, including when the laser fields are frequency-shifted in a stationary plasma or when the plasma flows at the sound speed in the direction of the laser propagation.

Laser smoothing techniques, such as Random Phase Plates (RPP) and Polarization Smoothing (PS), reduce LPI by degrading the spatio-temporal coherence of the laser. Smoothing by spectral dispersion (SSD) further broadens the temporal spectrum of the laser and disperses spatially the different frequencies across the beam aperture \cite{Boehly_1999, Skupski_1989, Garnier_1997, Garnier_2001}. These techniques lead to the formation of micron-scale speckles that evolve over picosecond time scales but are usually ignored for the design (and interpretation) of ICF experiments \cite{Strozzi2017,Kritcher2022,Higginson_2022,Liberatore_2023}, only accounting for PS. However, the micro-structure of the beams significantly alters the CBET dynamics as shown by recent studies which have improved our understanding of CBET  \cite{Oudin2021,Oudin2022,Seaton2022-2,Follett_2023}. For example, in the case of two RPP beams, the flow-driven grating leads to IAWs confined within the speckle region, while the frequency-shifted case allows the IAWs to propagate beyond the speckle, reducing the maximum power transfer and increasing the resonance width \cite{Oudin2025}. 

We use a  formalism that, for the first time, fully accounts for both spatial phase plate and temporal spectral dispersion smoothing effects on CBET in realistic  geometries. Our model, detailed in the companion paper Ref. \cite{lalaire}, reveals previously unsuspected key parameters that dramatically influence power transfer dynamics. Unlike what is usually assumed,  we show that drift velocity components normal to the ion acoustic wave direction and the finite coherence time of laser speckles fundamentally alter the wave mixing process, resulting in a resonance broadening beyond plane wave model predictions. This letter uses the theoretical framework detailed in Ref. \cite{lalaire} to  identify the critical parameters affecting CBET. We provide simple quantitative criteria based on the resonance width to determine when optical smoothing effects become dominant, with   profound implications for interpreting existing experiments and designing future fusion facilities.

%
The laser modeled here includes smoothing techniques relevant to ICF-relevant facilities. We consider both three dimensional (3D, $\mathcal{D}=2$) and two dimensional (2D, $\mathcal{D}=1$)  geometries, where the beam first passes through a random phase plate (RPP) composed of \( N^\mathcal{D} \) elements, which phase-shift the wavelets randomly. In 3D, these elements are indexed in the transverse \( y \) and \( z \) directions, by \( \mathbf{n} = (n_y, n_z) \), where \( n_y \) and \( n_z \) are integers between \( 1 \) and \( N \). In 2D, $\mathbf{n} \equiv n_y$.
In addition, the beam is smoothed by SSD. 
%
%
%
Regarding SSD, the frequency spectrum of the beam is modulated  with a frequency $\omega_d$ and a modulation depth $M$. 
This smoothing  introduces an effective bandwidth \( 2M \omega_d \) \cite{Skupsky1989}.
The beam has a spatial envelope of size $L$, a  time origin of the SSD frequency modulation  \(t_0\) and a  polarization vector  $\mathbf{u}$.

\begin{figure*}
    \centering
    \begin{tabular}{ccc} 
    (a) $\mathbf{v}_d=v_{dx}\Hat{\mathbf{x}}+v_{dy}\Hat{\mathbf{y}} $,
    $\omega=0$ & (b) $\mathbf{v}_d=v_{dy}\Hat{\mathbf{y}}$,
    $\omega=0$ &  (c) $\mathbf{v}_d=0 $,
    $\omega\neq 0$ \\
    \includegraphics[width = 0.3\linewidth]{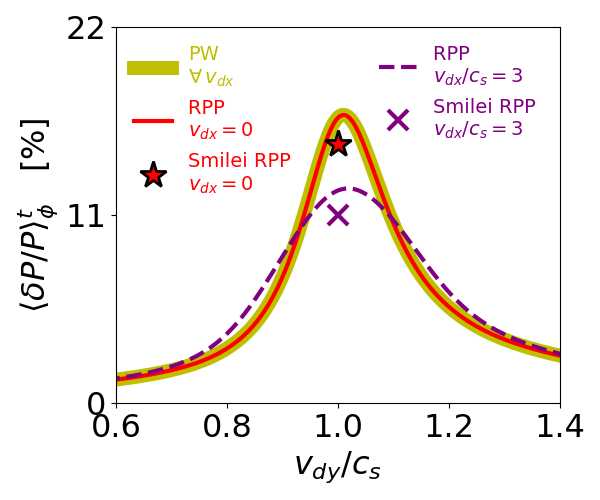}&
    \includegraphics[width = 0.3\linewidth]{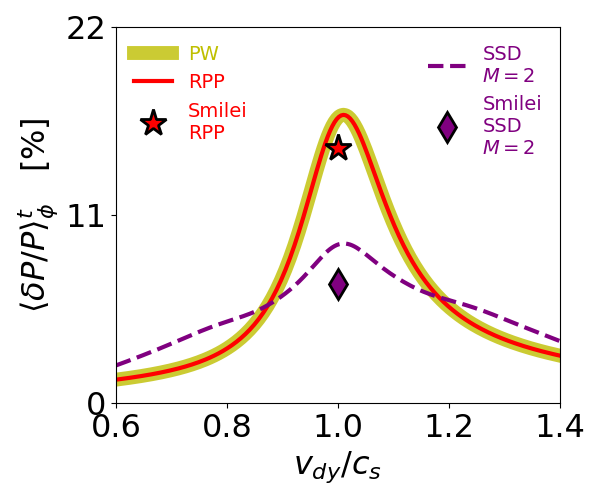}&
    \includegraphics[width = 0.3\linewidth]{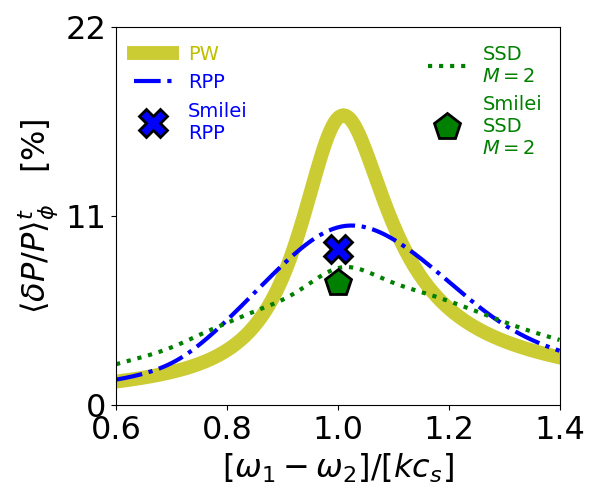}
    \end{tabular}
    \caption{ 
    %
    Power exchange predictions (lines, Eq. \eqref{eq:dPssd}) vs plasma conditions: (a,b) drift velocity $v_{dy}/c_s$ with $\omega=0$, (c) frequency shift $\omega/kc_s$ with $\mathbf{v}_d=0$. Simulation parameters: $n_e=0.04n_c$, $T_e=2$ keV, $T_i=1$ keV, $f_\#=8$, $\lambda_0=1\,\rm \mu m$, $I_0=40\,\rm TW/cm^2$ and $\omega_d/ 2\pi =   14.25~\text{GHz}$. PIC simulation results shown as markers \cite{lalaire}.  
    \label{fig:dp}
    }
\end{figure*}
The crossing of two beams induces a ponderomotive grating that drives an acoustic wave on which the light scatters. 
The $x$, $y$ and $z$ axis are set so that the acoustic propagation direction is $y$ and the $z$ is normal to the crossing plane. 
Assuming the  paraxial propagation of two laser beams of pulsation and wavevector $\omega_{1,2}$ and $\mathbf{k}_{1,2}$, we first write their ponderomotive beating of pulsation $\omega = \omega_1-\omega_2$ and wavevector $\mathbf{k} =\mathbf{k}_{1}-\mathbf{k}_2\simeq 2k_0\sin\theta\Hat{\mathbf{y}}$ (where $\theta$ is the crossing angle). We deduce the driven density fluctuations that derives from a linearized (fluid or kinetic) plasma response. We then derive the power exchange, also in the linearized framework,  such that the 
powers 
of lasers 1 and 2 may be written as 
$P_{1/2} =P_{1/2}^0+\delta P_{1/2}$
where 
$\delta P_{1/2}\ll P_{1/2}^0$.
We thus neglect the pump depletion and other non-linear effects such as particle trapping. 
 Following Ref. \cite{Oudin2025}, we then obtain a local value of the intensity exchange which needs to be integrated over the crossing region of extent $L$ in order to obtain the power exchange, $\delta P = \delta P_1^0-\delta P_2^0$. 
The two last simplification steps consist in averaging the power exchange on the phase plate variable, owing to the fact that the associated statistical variability is negligible for large enough phase plate element number, as shown in \cite{Oudin2025} in two dimensions.    
We  also restrict our analysis to temporal average   over a modulation period $2\pi/\omega_d$ of the power exchange and leading to,
\begin{align}
    \left\langle\frac{\delta P}{ P }\right\rangle_t= &-  \frac{CI_2^0}{N^{2\mathcal{D}}} \frac{L}{\sin{2 \theta}}   \sum_{\mathbf{n}_1,\mathbf{n}_2} \sum_{m_1,m_2,m_3 }  J_{m_1}J_{m_2}J_{m_3}
    \nonumber \\ 
           &\times J_{m_3+m_2-m_1}{\rm e}^{  -i(m_1-m_3)\omega_d t_0}    
    \nonumber \\     &\times  
    {\rm e}^{ +i\Psi_{1,m_1,\mathbf{n}_1}- i\Psi_{1,m_3,\mathbf{n}_1}-i\Psi_{2,m_1,\mathbf{n}_2}+ i\Psi_{2,m_3,\mathbf{n}_2} } 
    \nonumber \\     &\times 
   f_K[\omega+(m_1-m_2)\omega_d,\textbf{K}_{\textbf{n}_{1}\textbf{n}_{2}}] \, ,
   \label{eq:dPssd} \\
    C =& 
    (Z e^2 \omega_{pe}^2(\mathbf{u}_1\cdot \mathbf{u}_2)^2)/(2 m_i m_e \epsilon_0 k_0 \omega_0^2 c^2 c_s^2 v_g)
        \, , \nonumber  \\
    \textbf{K}_{\textbf{n}_{1}\textbf{n}_{2}}&= \big\{\sin(\theta)[k_{n_{y_1}}+k_{n_{y_2}}]\big\} \hat{\textbf{x}}+\nonumber\\&\big\{k-\cos(\theta)[k_{n_{y_1}}-k_{n_{y_2}}]\big\}\hat{\textbf{y}} + \big\{k_{n_{z_1}}-k_{n_{z_2}}\big\}\hat{\textbf{z}} \, , \nonumber
\end{align} 
where $\epsilon_0 $, $v_g$  and $\omega_{pe}$ are the electric permittivity, the laser group velocity and the electron plasma frequency, respectively.
We assumed $P_1^0=P_2^0$ for simplicity and introduced $P = P_1^0+P_2^0$, the sound speed $c_s$ and $m_{e/i}$ and $c$, the electron/ion mass and the light speed in vacuum.
The two sums over \( \mathbf{n} \) correspond to the two phase plates of the crossing lasers. 
%
%
%
The wavevectors \( \mathbf{k}_\mathbf{n} = (0,k_{n_y}, k_{n_z}) \) are distributed according to the geometry of the beams and depend on the number of elements \( N \), the f-number \( f_\# \).
The two discrete sums over \( m \)  stem from the frequency modulation of the two SSD 
where \( J_m(M)\) is the Bessel function of the first kind.  The phase \( \Psi_{b,m,\mathbf{n}} \) characterizes the spatial dispersion of the different laser frequencies and  whose specific form depends on the smoothing technique that is used \cite{lalaire}. 

The plasma response function is here kinetic,
    $f_K(\Omega,\mathbf{K})=  \mathbf{K}^2  \lambda_{De}^2 \chi_e
    (1+\sum_i\chi_i)/(1+\sum_i\chi_i+\chi_e)$
where the subscript $e$ and $i$ designate the electrons and the different ion populations  which compose the plasma. We also introduced the electrostatic susceptibilities 
$\chi_{e/i}$ \cite{Fried_1960}, derived at a wavevector $ \textbf{K}_{\textbf{n}_{1}\textbf{n}_{2}}$ and pulsation $\Omega_{m_1,m_2}=\omega+(m_1-m_2)\omega_d$.
Note that this power exchange, although averaged on a SSD modulation period, does depend on the  time delay between the two SSD frequency modulators, $t_0$. 
Accounting for PS simply consists in replacing $(\mathbf{u}_1\cdot \mathbf{u}_2)^2$ by $1/2$ \cite{Michel2009b}. 
The sums in this expression show that the power transfer results from the superposition of  the different interactions between the beating electromagnetic wavelets with the driven acoustic wavelets. 

This expression is  simplified  by an average over $t_0\in[0 ,2\pi/\omega_d]$, giving,
\begin{align}
    \frac{\langle \delta P\rangle_\phi^{t,t_0}}{P}
    &= -\frac{P_2^0\,C}{\sin(2\theta)\, N^{2\mathcal{D}}}
    \sum_{\mathbf{n}_1}\sum_{\mathbf{n}_2}\sum_{m_1}\sum_{m_2}
    J_{m_1}^2(M)\,J_{m_2}^2(M)\, \nonumber \\ 
    &\times \Im\!\left[f_F\!\left(\bar{v}_{\phi_{\mathbf{n}_1,\mathbf{n}_2,m_1,m_2}}\right)\right]\, .
    \label{eq:dPssd_avet0}
\end{align}
This step removes the dependence of the power transfer on the spatial frequency dispersion $\Psi_{b,m,\mathbf{n}}$. Such simplification leads to relative deviations of the predicted power transfer up to $\sim 40\%$,  especially in low Landau damped plasmas, as shown in the companion paper Ref. \cite{lalaire}. However, the simplicity of the resulting formula allows to illustrate and analyze the different mechanisms responsible of the deviation of the power transfer between smoothed  beams from the plane wave predictions.  
Similarly, we here use a fluid plasma response,  
$f_F(\bar{v}_\phi) \approx 
m_e\,\omega_{pe}^2/(m_i\,c_s^2)/ (1-\bar{v}_\phi^2 - 2 i\,\Bar{\nu_{\mathrm{L}}}\,\bar{v}_\phi)$.
We introduced the normalized phase velocity of the driven IAWs,
$\bar{v}_{\phi_{\mathbf{n}_1,\mathbf{n}_2,m_1,m_2}}
= (\Omega_{m_1,m_2} - \mathbf{K}_{\mathbf{n}_1,\mathbf{n}_2}\!\cdot\mathbf{v}_d)/ (
       |\mathbf{K}_{\mathbf{n}_1,\mathbf{n}_2}|\,c_s)$
and  
\(\Bar{\nu_{\mathrm{L}}}=\nu_L/\vert K_{\mathbf{n}_1,\mathbf{n}_2}\vert c_s\), 
the normalized Landau damping rate.  

The comparison of our model with a kinetic plasma response [Eq. \eqref{eq:dPssd}] is made in light of Smilei \cite{Derouillat2018} particle-in-cell (PIC) simulations and is detailed in  Ref. \cite{lalaire}. 
The plasma  is composed of C$^{6+}$ and H$^{+}$ ions of equal proportion with a homogeneous density and temperature $n_e=0.04n_c$ (where $n_c\simeq 10^{21}\,\rm cm^{-3}$), $T_e=2\,\rm keV$ and $T_i=1\,\rm keV$. The mean intensity of both P-polarized beams is  $I_0=40\,\rm TW/cm^2$ with a waist of $L_\perp = 48\,\rm \mu m$ and $t_0=0$ for a central laser wavelength of $1\rm\mu m$. We performed longitudinal  SSD corresponding to $t_0=0$, $\omega_d/ 2\pi =   14.25~\text{GHz}$ and  $\Psi_{b,m,\mathbf{n}}= f_0 (\omega_d/\omega_0) \mathbf{k}_\mathbf{n}^2/2k_0$ where $f_0=8\,\rm m$ is the focal length.
Other numerical details can be found in Ref. \cite{lalaire}. 
Note that we validated the model with  paraxial hydrodynamic Hera simulations \cite{Loiseau_2006} as detailed in Ref. \cite{lalaire}.

When plotted against  the resonance parameter $(\omega-kv_{dy})/(kc_s)$, the power transfer exhibits a peak of width which depends on the other parameters of the system.  We now illustrate the CBET between smoothed beams in three representative situations and examine the dependence of the resonance peak on the flow and optical parameters.  
The first situation is illustrated in Fig. \ref{fig:dp}(a) and corresponds to spatially smoothed beams with equal wavelength ($\omega=0)$, without temporal smoothing and crossing in a plasma flowing with velocity $\mathbf{v}_d=v_{dx}\Hat{\mathbf{x}}+v_{dy}\Hat{\mathbf{y}}$. When $v_{dx}=0$, the RPP curve as a  red solid line coincides with the plane wave (PW) limit as a yellow solid line. Indeed, the IAW do not leave the speckle vicinity \cite{Oudin2025}, resulting in a resonance curve of maximum inversely proportional to the damping of the IAW, $\nu_L$ and of full width at half maximum (FWHM) $\sigma_{L}  = 2\nu_L/kc_s$ (which is here around $\nu_L/kc_s\simeq 0.1$).  The flow component $v_{dx}=3c_s$  advects the IAW away from the hot spot regions, thus broadening the resonance and decreasing its maximum, as shown by the purple dashed line.
The new resonance width has an additional contribution that can be extracted from our theory. For that,  the large phase plate element number limit allows to replace the discrete sums of Eq. \eqref{eq:dPssd_avet0} by continuous ones. We also write Eq. \eqref{eq:dPssd_avet0} for a vanishing IAW damping ($\Bar{\nu_L}\to 0$), thus using $\Im\{f_F(\bar{v}_\phi)\} \approx (\pi/2)\,\delta(\bar{v}_\phi - 1)$ (where $\delta(x)$ is the Dirac distribution). Finally, assuming $f_\#\sin\theta\gg 1$ leads to $\bar{v}_\phi(\Delta k_X) \approx (v_{dy}/c_s) + \sin(\theta)\,(v_{dx}/c_s)\,\Delta k_X/k$ where \(\Delta k_{X} = k_{n_{y_1}} + k_{n_{y_2}} \in [-2\tilde{k},2\tilde{k}]\). The resulting expression shows explicitly a width of  $\sigma_{v_{dx}} = \vert v_{dx}\vert /2f_\# c_s $ (see the supplemental material \cite{supp}).
For a finite value of the Landau damping, 
we observe that the total resonance width results from a  quadratic sum between the Landau damping and the flow contributions, following $\sigma^2 \simeq  \sigma_{L}^2 + \sigma_{v_{dx}}^2$. 
The markers in Fig. \ref{fig:dp}(a) correspond to PIC results and confirm our predictions.  Likewise,  a $v_{dz}$ flow component further broadens the resonance through the contribution $\sigma_{v_{dz}} = \vert v_{dz}/[2c_sf_\# \sin(\theta)]\vert$.

The second physical situation corresponds to $v_{dx}=v_{dz}=0$ and temporally smoothed laser beams crossing with identical central wavelengths. The effect of temporal smoothing is illustrated by the purple dashed line in Fig. \ref{fig:dp}(b) and has, as the previous case, a smaller maximum and a larger width than the plane wave predictions (as a yellow line). Here, the resonance curve has a width related to the laser bandwidth, $2M\omega_d$ and which is also observed to add quadratically to the plane wave limit according to $\sigma^2 \simeq  \sigma_{L}^2 + \sigma_{SSD}^2$. The evaluation of $\sigma_{SSD}$ requires, as in the previous case, the simplification of Eq. \eqref{eq:dPssd_avet0} in the low Landau damping and large phase plate element number limits. The laser Bessel temporal spectrum is  here replaced by a flat one for simplicity,  with a similar procedure as in the previous case  \cite{supp}, we obtain $ \sigma_{SSD}\simeq 2M\omega_d/kc_s$. Hence, the speckle motion imposed by the SSD effectively smooths the driven density fluctuations, thus decreasing the maximum power transfer, in agreement with the PIC simulations (as markers). Note that our expression of $\sigma_{SSD}$ remains valid while $2M\omega_d\lesssim kc_s$ \cite{supp}.

Figure \ref{fig:dp}(c) shows the crossing of temporally and spatially smoothed beams with a frequency shift in a plasma at rest ($\mathbf{v}_d=0)$. Without SSD [see the blue curve], the resonance is broadened by the frequency shift. The ponderomotive grating propagates in the laboratory frame, allowing the driven IAW to leave the crossing speckle region.  In Ref. \cite{Oudin2025}, assuming a low damping rate, we demonstrate that the resonance width fulfills $\sigma_{ \omega\neq 0} = \delta_{\omega_1,\omega_2}/2\vert \tan(\theta)\vert f_\#$ (where $\delta_{\omega_1,\omega_2}$ is the Kronecker symbol). Hence, the blue line of Fig. \ref{fig:dp}(c) has a width given by   $\sigma^2 \simeq  \sigma_{L}^2 + \sigma_{\omega\neq 0}^2$. Adding SSD to the two beams further broadens the resonance as illustrated by the green dotted line and leads to  $\sigma^2 \simeq  \sigma_{L}^2 +\sigma_{\omega\neq 0}^2+ \sigma_{SSD}^2$. Once again, the power transfer predictions are confirmed by the PIC data as markers. 

\begin{figure}
    \centering
    \begin{tabular}{c} 
    (a)  $\omega\neq 0$, $v_{dx}=0$   \\
    \includegraphics[width = 0.7\linewidth]{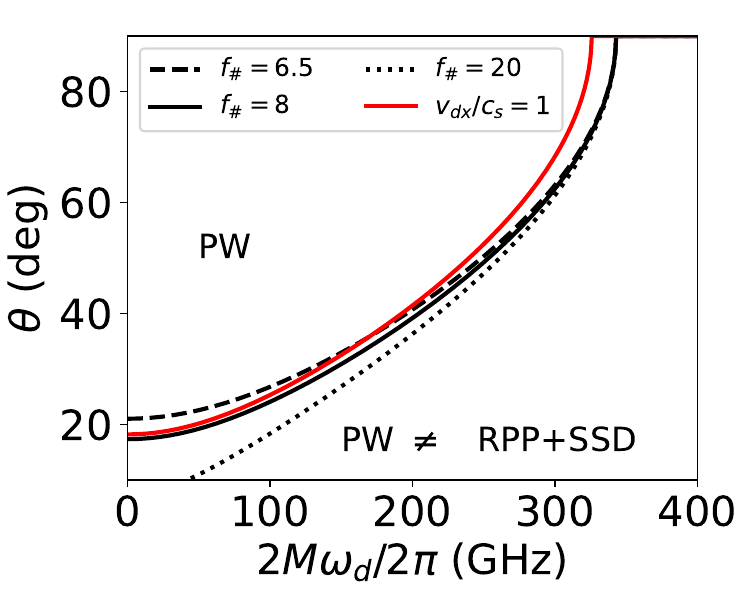}\\
    (b)      $\omega=0$ \\
    \includegraphics[width = 0.7\linewidth]{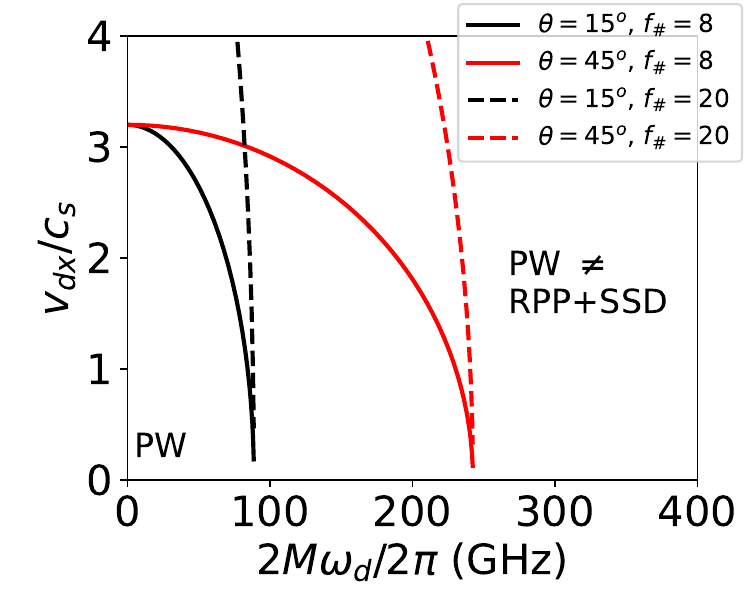}
    \end{tabular}
    \caption{ Threshold [Eq. \eqref{eq:sigma}] for a plasma flow $\mathbf{v}=v_{dx}\Hat{\mathbf{x}}+v_{dy}\Hat{\mathbf{y}} $ ($v_{dz}=\sigma_{v_{dz}}=0$) for a frequency shift (a) and in the drift case (b). The region where plane wave CBET models are (not) satisfactory is noted "PW" ("PW$\neq$RPP+SSD"). (a) Black lines: $v_{dx}=0$ for various f-number; solid red line: $v_{dx}=c_s$ and $f_\#=8$. The lasers have   $\lambda_0=0.35\,\rm \mu m$, we set $c_s=10^{-3}c$ and $\nu_L/kc_s=0.1$.
    %
    \label{fig:param}
    }
\end{figure}
We demonstrate the spatial smoothing does not impact the  CBET resonance when $ \sigma_{\omega\neq 0}\ll \sigma_{L}$, specifically, when $\tan\theta \gg 4f_\# \nu_L/kc_s$. This is consistent with the CBET measurements performed at large crossing angle that are well reproduced by the plane wave theory \cite{Neuville_2016,Neuville2018,Turnbull_2020,Hansen2021}. Reference \cite{Debayle_2025} presents CBET measurements performed at small angle, but suffers from a target misalignment preventing any quantitative evidence of the impact of optical smoothing.  When accounting for all the physical situations addressed in this letter, SSD, $v_{dx}\neq 0$ and $v_{dz}\neq 0$, we find that laser smoothing modifies the power transfer compared to the plane wave limit when the width of the resonance deviates from the Landau damping. The corresponding threshold which discriminates a case where the power transfer is well reproduced within the plane wave framework from a case which is not is thus, \begin{align}\label{eq:sigma}
\sigma_{\omega\neq 0}^2+ \sigma_{SSD}^2+\sigma_{v_{dx}}^2 +\sigma_{v_{dz}}^2  = \sigma_{L}^2\, .
\end{align}
When the plasma drifts only in the $y$ direction, this expression gives the crossing angle as a function of the laser bandwidth depending on the laser f-number, as illustrated by the black lines in Fig. \ref{fig:param}(a) for $3\omega$ lasers and a large Landau damping $\nu_L/kc_s=0.1$. This shows that without SSD ($M\omega_d=0$), CBET at half crossing angle above $\sim 20^o $ is well described by the plane wave limit ("PW"), as shown in Ref. \cite{Oudin2025}. When adding SSD, the region of parameters where plane wave models are satisfactory shrinks and vanishes for bandwidth above $\sim 300\, \rm GHz$. As a comparison, the NIF frequency bandwidth at $3\omega$  used on the N210808 NIF shot \cite{PRE_Kritcher_2022} is $\sim 135 \, \rm GHz$. For LMJ and Omega, the nominal value is $430$ and $360\,\rm GHz$, respectively. As expected, the validity of the  plane wave model is larger for large aperture/large speckles ($f_\#=20$ as a dotted black line) than for small aperture beams ($f_\#=6.5$ as dashed line). The red solid line considers a plasma drifting with a $v_{dx}$-component and with $f_\#=8$, it highlights a significant sensitivity of this threshold to the flow direction. The dependence of  the CBET model on the $x$ flow component is illustrated in  Fig. \ref{fig:param}(b)  without frequency shift between the lasers and  as a function of the laser bandwidth.  Only low $v_{dx}$ velocities and low bandwidth are well described by a plane wave CBET model. Laser bandwidth above $\sim 100\, \rm GHz$ at $\theta =15^o$ ($\sim 250\, \rm GHz$ at $\theta =45^o$) or flows $v_{dx}\gtrsim 3c_s$ require a CBET model with smoothing ("PW$\neq$RPP+SSD").  Here again, the range of validity of the plane wave models is larger for large than for small aperture beams. 

Regarding the influence of the $v_{dz}$ component $\sigma_{v_{dz}} = \sigma_{L}$ leads to $\vert v_{dz}\vert =4\Bar{\nu_L} f_\#c_s\vert \sin(\theta)\vert $. Consequently, a value of $ v_{dz}$ of the order of the sound speed induces significant discrepancies on the power transfer compared to the PW approximation. This effect may be negligible for large enough laser bandwidth, when $\sigma_{SSD} >\sigma_{v_{dz}} $, giving  $2M\omega_d/2\pi \gtrsim k_0\vert v_{dz}\vert /f_\# \simeq   240 (\vert v_{dz}\vert /c_s)\, \rm GHz$ (for $c_s=10^{-3}c$ and $f_\#=8$). To date, this condition  for $v_{dz}=2c_s$ is not fulfilled in any high energy laser facility. For a facility such as NIF, $v_{dz}\simeq c_s/2$ is enough to perturb the CBET. 

\begin{figure}
    \centering
    \begin{tabular}{cc} 
    (a)  Plane wave &(b)    RPP+SSD \\
    \includegraphics[width = 0.24\textwidth]{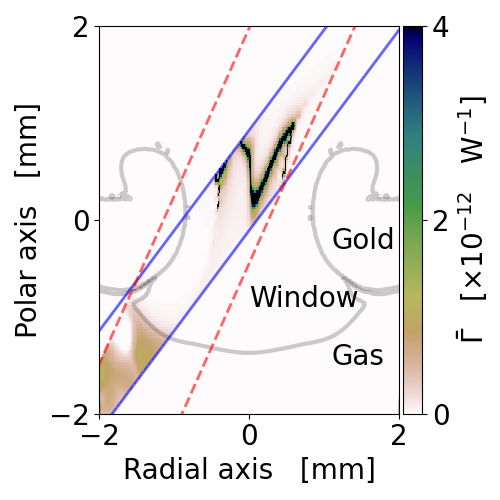} & 
    \includegraphics[width = 0.24\textwidth]{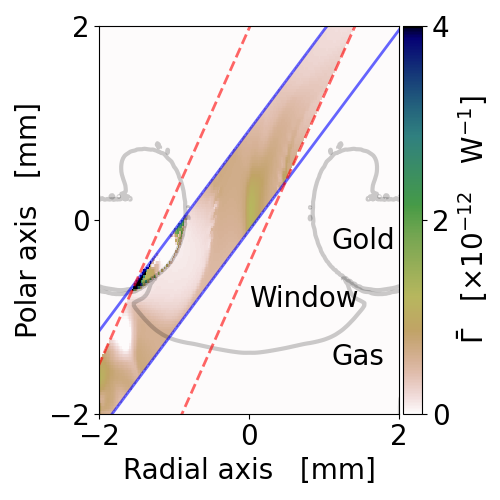}
    \end{tabular}
    \caption{ CBET coupling parameter calculated at the entrance hall of  a Troll radiative hydrodynamic simulation of the N210808 NIF shot at $6\,\rm ns$ in the plane wave (a) and smoothed beam (b) cases. The material boundaries are the solid grey lines. The $30^o$ and $44^o$ cone edges are superimposed as dashed red and blue solid lines.  
    \label{fig:nif}
    }
\end{figure} 
Our theoretical model has been applied to analyze and predict the outcomes of specific experimental setups at the National Ignition Facility (NIF). We performed a detailed Troll radiative hydrodynamic 2D axi-symmetric simulation \cite{Lefebvre_2018} of the NIF high-yield shot N210808 \cite{PRL_Abu-Shawareb_2022} following Refs. \cite{Kritcher2022,PRE_Kritcher_2022}. The CBET coupling parameter (defined here as $\Bar{\Gamma}= \left\langle\delta P/P \right\rangle_t \sin(2\theta)/ (LP_2^0)$ combined with Eq. \eqref{eq:dPssd_avet0}) with the specific NIF laser parameters, is illustrated in Fig.~\ref{fig:nif} at 6 ns during the main laser drive in the crossing region of the 30$^\circ$ and 44$^\circ$ cones.
Regarding the laser entrance region, the usual plane wave case presented in panel (a) shows sharp structures located at the CBET resonance. Indeed, the resonance width is dominated here by the local value of the Landau damping frequency. When adding the influence of the optical smoothing, panel (b) presents much broader structures with a resonance width primarily related to the laser bandwidth but also to the laser aperture and flow component normal to the IAWs direction. In this case, the power exchange should be much more gradual than for the crossing of two plane waves, thus modifying the intensity profile of the lasers in the hohlraum.

We have demonstrated the importance of accounting for the influence of temporal smoothing and flow components normal to the IAW direction on the CBET dynamics, under realistic conditions. 
Our analytical model, based on a linear approach and assuming paraxial approximation detailed in the companion paper Ref.  \cite{lalaire}, predicts  the power transfer depending on the laser smoothing and plasma parameters, which would not be possible with a plane wave model.
The use of SSD leads to a broader resonance width and a reduction in the maximum power transfer. Likewise, a misalignment of the flow with the IAW also results in a broader resonance. We derived simple analytical estimates that allow to pinpoint which laser and plasma parameters require supplementing CBET models with laser smoothing. This suggests  that SSD above $\sim 200\,\rm GHz$ or $x$-flow component above $\sim c_s$ results in a different power transfer than the usual plane wave predictions depending on the crossing angle and beams aperture.  Moreover, the power transfer between smoothed beams is sensitive to  a $z$-aligned drift velocity component that can arise due to irradiation symmetry defects.
In a realistic non-homogeneous plasma, the  power transfer resulting from a broader resonance  will occur more gradually with optical smoothing than without, significantly changing the intensity profiles and subsequent backscattering and energy deposition properties. 
%
Our model also underlines the significant  impact of the two SSD modulator synchronization on the CBET outcome, as detailed in Ref. \cite{lalaire}. 
By accounting for spatial and temporal smoothing, we can more accurately predict and control CBET, paving the way for improved experimental designs and outcomes in ICF research.

This work has been done under the auspices of Commissariat à l’Energie Atomique (CEA), and the simulations were performed using high performance computing resources at Centre de Calcul pour la Recherche et la Technologie and CEA/Tera.

\bibliographystyle{aip}
\bibliography{biblio}

@article{PRL_Fusion_2022,
  title = {Lawson Criterion for Ignition Exceeded in an Inertial Fusion Experiment},
  author = {Abu-Shawareb, H. and Acree, R. and Adams, P. and Adams, J. and Addis, B. and Aden, R. and Adrian, P. and Afeyan, B. B. and Aggleton, M. and Aghaian, L. and Aguirre, A. and Aikens, D. and Akre, J. and Albert, F. and Albrecht, M. and Albright, B. J. and Albritton, J. and Alcala, J. and Alday, C. and Alessi, D. A. and Alexander, N. and Alfonso, J. and Alfonso, N. and Alger, E. and Ali, S. J. and Ali, Z. A. and Alley, W. E. and Amala, P. and Amendt, P. A. and Amick, P. and Ammula, S. and Amorin, C. and Ampleford, D. J. and Anderson, R. W. and Anklam, T. and Antipa, N. and Appelbe, B. and Aracne-Ruddle, C. and Araya, E. and Arend, M. and Arnold, P. and Arnold, T. and Asay, J. and Atherton, L. J. and Atkinson, D. and Atkinson, R. and Auerbach, J. M. and Austin, B. and Auyang, L. and Awwal, A. S. and Ayers, J. and Ayers, S. and Ayers, T. and Azevedo, S. and Bachmann, B. and Back, C. A. and Bae, J. and Bailey, D. S. and Bailey, J. and Baisden, T. and Baker, K. L. and Baldis, H. and Barber, D. and Barberis, M. and Barker, D. and Barnes, A. and Barnes, C. W. and Barrios, M. A. and Barty, C. and Bass, I. and Batha, S. H. and Baxamusa, S. H. and Bazan, G. and Beagle, J. K. and Beale, R. and Beck, B. R. and Beck, J. B. and Bedzyk, M. and Beeler, R. G. and Beeler, R. G. and Behrendt, W. and Belk, L. and Bell, P. and Belyaev, M. and Benage, J. F. and Bennett, G. and Benedetti, L. R. and Benedict, L. X. and Berger, R. and Bernat, T. and Bernstein, L. A. and Berry, B. and Bertolini, L. and Besenbruch, G. and Betcher, J. and Bettenhausen, R. and Betti, R. and Bezzerides, B. and Bhandarkar, S. D. and Bickel, R. and Biener, J. and Biesiada, T. and Bigelow, K. and Bigelow-Granillo, J. and Bigman, V. and Bionta, R. M. and Birge, N. W. and Bitter, M. and Black, A. C. and Bleile, R. and Bleuel, D. L. and Bliss, E. and Bliss, E. and Blue, B. and Boehly, T. and Boehm, K. and Boley, C. D. and Bonanno, R. and Bond, E. J. and Bond, T. and Bonino, M. J. and Borden, M. and Bourgade, J.-L. and Bousquet, J. and Bowers, J. and Bowers, M. and Boyd, R. and Bozek, A. and Bradley, D. K. and Bradley, K. S. and Bradley, P. A. and Bradley, L. and Brannon, L. and Brantley, P. S. and Braun, D. and Braun, T. and Brienza-Larsen, K. and Briggs, T. M. and Britten, J. and Brooks, E. D. and Browning, D. and Bruhn, M. W. and Brunner, T. A. and Bruns, H. and Brunton, G. and Bryant, B. and Buczek, T. and Bude, J. and Buitano, L. and Burkhart, S. and Burmark, J. and Burnham, A. and Burr, R. and Busby, L. E. and Butlin, B. and Cabeltis, R. and Cable, M. and Cabot, W. H. and Cagadas, B. and Caggiano, J. and Cahayag, R. and Caldwell, S. E. and Calkins, S. and Callahan, D. A. and Calleja-Aguirre, J. and Camara, L. and Camp, D. and Campbell, E. M. and Campbell, J. H. and Carey, B. and Carey, R. and Carlisle, K. and Carlson, L. and Carman, L. and Carmichael, J. and Carpenter, A. and Carr, C. and Carrera, J. A. and Casavant, D. and Casey, A. and Casey, D. T. and Castillo, A. and Castillo, E. and Castor, J. I. and Castro, C. and Caughey, W. and Cavitt, R. and Celeste, J. and Celliers, P. M. and Cerjan, C. and Chandler, G. and Chang, B. and Chang, C. and Chang, J. and Chang, L. and Chapman, R. and Chapman, T. and Chase, L. and Chen, H. and Chen, H. and Chen, K. and Chen, L.-Y. and Cheng, B. and Chittenden, J. and Choate, C. and Chou, J. and Chrien, R. E. and Chrisp, M. and Christensen, K. and Christensen, M. and Christopherson, A. R. and Chung, M. and Church, J. A. and Clark, A. and Clark, D. S. and Clark, K. and Clark, R. and Claus, L. and Cline, B. and Cline, J. A. and Cobble, J. A. and Cochrane, K. and Cohen, B. and Cohen, S. and Collette, M. R. and Collins, G. and Collins, L. A. and Collins, T. J. B. and Conder, A. and Conrad, B. and Conyers, M. and Cook, A. W. and Cook, D. and Cook, R. and Cooley, J. C. and Cooper, G. and Cope, T. and Copeland, S. R. and Coppari, F. and Cortez, J. and Cox, J. and Crandall, D. H. and Crane, J. and Craxton, R. S. and Cray, M. and Crilly, A. and Crippen, J. W. and Cross, D. and Cuneo, M. and Cuotts, G. and Czajka, C. E. and Czechowicz, D. and Daly, T. and Danforth, P. and Darbee, R. and Darlington, B. and Datte, P. and Dauffy, L. and Davalos, G. and Davidovits, S. and Davis, P. and Davis, J. and Dawson, S. and Day, R. D. and Day, T. H. and Dayton, M. and Deck, C. and Decker, C. and Deeney, C. and DeFriend, K. A. and Deis, G. and Delamater, N. D. and Delettrez, J. A. and Demaret, R. and Demos, S. and Dempsey, S. M. and Desjardin, R. and Desjardins, T. and Desjarlais, M. P. and Dewald, E. L. and DeYoreo, J. and Diaz, S. and Dimonte, G. and Dittrich, T. R. and Divol, L. and Dixit, S. N. and Dixon, J. and Dodd, E. S. and Dolan, D. and Donovan, A. and Donovan, M. and D\"oppner, T. and Dorrer, C. and Dorsano, N. and Douglas, M. R. and Dow, D. and Downie, J. and Downing, E. and Dozieres, M. and Draggoo, V. and Drake, D. and Drake, R. P. and Drake, T. and Dreifuerst, G. and DuBois, D. F. and DuBois, P. F. and Dunham, G. and Dylla-Spears, R. and Dymoke-Bradshaw, A. K. L. and Dzenitis, B. and Ebbers, C. and Eckart, M. and Eddinger, S. and Eder, D. and Edgell, D. and Edwards, M. J. and Efthimion, P. and Eggert, J. H. and Ehrlich, B. and Ehrmann, P. and Elhadj, S. and Ellerbee, C. and Elliott, N. S. and Ellison, C. L. and Elsner, F. and Emerich, M. and Engelhorn, K. and England, T. and English, E. and Epperson, P. and Epstein, R. and Erbert, G. and Erickson, M. A. and Erskine, D. J. and Erlandson, A. and Espinosa, R. J. and Estes, C. and Estabrook, K. G. and Evans, S. and Fabyan, A. and Fair, J. and Fallejo, R. and Farmer, N. and Farmer, W. A. and Farrell, M. and Fatherley, V. E. and Fedorov, M. and Feigenbaum, E. and Feit, M. and Ferguson, W. and Fernandez, J. C. and Fernandez-Panella, A. and Fess, S. and Field, J. E. and Filip, C. V. and Fincke, J. R. and Finn, T. and Finnegan, S. M. and Finucane, R. G. and Fischer, M. and Fisher, A. and Fisher, J. and Fishler, B. and Fittinghoff, D. and Fitzsimmons, P. and Flegel, M. and Flippo, K. A. and Florio, J. and Folta, J. and Folta, P. and Foreman, L. R. and Forrest, C. and Forsman, A. and Fooks, J. and Foord, M. and Fortner, R. and Fournier, K. and Fratanduono, D. E. and Frazier, N. and Frazier, T. and Frederick, C. and Freeman, M. S. and Frenje, J. and Frey, D. and Frieders, G. and Friedrich, S. and Froula, D. H. and Fry, J. and Fuller, T. and Gaffney, J. and Gales, S. and Le Galloudec, B. and Le Galloudec, K. K. and Gambhir, A. and Gao, L. and Garbett, W. J. and Garcia, A. and Gates, C. and Gaut, E. and Gauthier, P. and Gavin, Z. and Gaylord, J. and Geissel, M. and G\'enin, F. and Georgeson, J. and Geppert-Kleinrath, H. and Geppert-Kleinrath, V. and Gharibyan, N. and Gibson, J. and Gibson, C. and Giraldez, E. and Glebov, V. and Glendinning, S. G. and Glenn, S. and Glenzer, S. H. and Goade, S. and Gobby, P. L. and Goldman, S. R. and Golick, B. and Gomez, M. and Goncharov, V. and Goodin, D. and Grabowski, P. and Grafil, E. and Graham, P. and Grandy, J. and Grasz, E. and Graziani, F. and Greenman, G. and Greenough, J. A. and Greenwood, A. and Gregori, G. and Green, T. and Griego, J. R. and Grim, G. P. and Grondalski, J. and Gross, S. and Guckian, J. and Guler, N. and Gunney, B. and Guss, G. and Haan, S. and Hackbarth, J. and Hackel, L. and Hackel, R. and Haefner, C. and Hagmann, C. and Hahn, K. D. and Hahn, S. and Haid, B. J. and Haines, B. M. and Hall, B. M. and Hall, C. and Hall, G. N. and Hamamoto, M. and Hamel, S. and Hamilton, C. E. and Hammel, B. A. and Hammer, J. H. and Hampton, G. and Hamza, A. and Handler, A. and Hansen, S. and Hanson, D. and Haque, R. and Harding, D. and Harding, E. and Hares, J. D. and Harris, D. B. and Harte, J. A. and Hartouni, E. P. and Hatarik, R. and Hatchett, S. and Hauer, A. A. and Havre, M. and Hawley, R. and Hayes, J. and Hayes, J. and Hayes, S. and Hayes-Sterbenz, A. and Haynam, C. A. and Haynes, D. A. and Headley, D. and Heal, A. and Heebner, J. E. and Heerey, S. and Heestand, G. M. and Heeter, R. and Hein, N. and Heinbockel, C. and Hendricks, C. and Henesian, M. and Heninger, J. and Henrikson, J. and Henry, E. A. and Herbold, E. B. and Hermann, M. R. and Hermes, G. and Hernandez, J. E. and Hernandez, V. J. and Herrmann, M. C. and Herrmann, H. W. and Herrera, O. D. and Hewett, D. and Hibbard, R. and Hicks, D. G. and Hill, D. and Hill, K. and Hilsabeck, T. and Hinkel, D. E. and Ho, D. D. and Ho, V. K. and Hoffer, J. K. and Hoffman, N. M. and Hohenberger, M. and Hohensee, M. and Hoke, W. and Holdener, D. and Holdener, F. and Holder, J. P. and Holko, B. and Holunga, D. and Holzrichter, J. F. and Honig, J. and Hoover, D. and Hopkins, D. and Berzak Hopkins, L. and Hoppe, M. and Hoppe, M. L. and Horner, J. and Hornung, R. and Horsfield, C. J. and Horvath, J. and Hotaling, D. and House, R. and Howell, L. and Hsing, W. W. and Hu, S. X. and Huang, H. and Huckins, J. and Hui, H. and Humbird, K. D. and Hund, J. and Hunt, J. and Hurricane, O. A. and Hutton, M. and Huynh, K. H.-K. and Inandan, L. and Iglesias, C. and Igumenshchev, I. V. and Izumi, N. and Jackson, M. and Jackson, J. and Jacobs, S. D. and James, G. and Jancaitis, K. and Jarboe, J. and Jarrott, L. C. and Jasion, D. and Jaquez, J. and Jeet, J. and Jenei, A. E. and Jensen, J. and Jimenez, J. and Jimenez, R. and Jobe, D. and Johal, Z. and Johns, H. M. and Johnson, D. and Johnson, M. A. and Gatu Johnson, M. and Johnson, R. J. and Johnson, S. and Johnson, S. A. and Johnson, T. and Jones, K. and Jones, O. and Jones, M. and Jorge, R. and Jorgenson, H. J. and Julian, M. and Jun, B. I. and Jungquist, R. and Kaae, J. and Kabadi, N. and Kaczala, D. and Kalantar, D. and Kangas, K. and Karasiev, V. V. and Karasik, M. and Karpenko, V. and Kasarky, A. and Kasper, K. and Kauffman, R. and Kaufman, M. I. and Keane, C. and Keaty, L. and Kegelmeyer, L. and Keiter, P. A. and Kellett, P. A. and Kellogg, J. and Kelly, J. H. and Kemic, S. and Kemp, A. J. and Kemp, G. E. and Kerbel, G. D. and Kershaw, D. and Kerr, S. M. and Kessler, T. J. and Key, M. H. and Khan, S. F. and Khater, H. and Kiikka, C. and Kilkenny, J. and Kim, Y. and Kim, Y.-J. and Kimko, J. and Kimmel, M. and Kindel, J. M. and King, J. and Kirkwood, R. K. and Klaus, L. and Klem, D. and Kline, J. L. and Klingmann, J. and Kluth, G. and Knapp, P. and Knauer, J. and Knipping, J. and Knudson, M. and Kobs, D. and Koch, J. and Kohut, T. and Kong, C. and Koning, J. M. and Koning, P. and Konior, S. and Kornblum, H. and Kot, L. B. and Kozioziemski, B. and Kozlowski, M. and Kozlowski, P. M. and Krammen, J. and Krasheninnikova, N. S. and Kraus, B. and Krauser, W. and Kress, J. D. and Kritcher, A. L. and Krieger, E. and Kroll, J. J. and Kruer, W. L. and Kruse, M. K. G. and Kucheyev, S. and Kumbera, M. and Kumpan, S. and Kunimune, J. and Kustowski, B. and Kwan, T. J. T. and Kyrala, G. A. and Laffite, S. and Lafon, M. and LaFortune, K. and Lahmann, B. and Lairson, B. and Landen, O. L. and Langenbrunner, J. and Lagin, L. and Land, T. and Lane, M. and Laney, D. and Langdon, A. B. and Langer, S. H. and Langro, A. and Lanier, N. E. and Lanier, T. E. and Larson, D. and Lasinski, B. F. and Lassle, D. and LaTray, D. and Lau, G. and Lau, N. and Laumann, C. and Laurence, A. and Laurence, T. A. and Lawson, J. and Le, H. P. and Leach, R. R. and Leal, L. and Leatherland, A. and LeChien, K. and Lechleiter, B. and Lee, A. and Lee, M. and Lee, T. and Leeper, R. J. and Lefebvre, E. and Leidinger, J.-P. and LeMire, B. and Lemke, R. W. and Lemos, N. C. and Le Pape, S. and Lerche, R. and Lerner, S. and Letts, S. and Levedahl, K. and Lewis, T. and Li, C. K. and Li, H. and Li, J. and Liao, W. and Liao, Z. M. and Liedahl, D. and Liebman, J. and Lindford, G. and Lindman, E. L. and Lindl, J. D. and Loey, H. and London, R. A. and Long, F. and Loomis, E. N. and Lopez, F. E. and Lopez, H. and Losbanos, E. and Loucks, S. and Lowe-Webb, R. and Lundgren, E. and Ludwigsen, A. P. and Luo, R. and Lusk, J. and Lyons, R. and Ma, T. and Macallop, Y. and MacDonald, M. J. and MacGowan, B. J. and Mack, J. M. and Mackinnon, A. J. and MacLaren, S. A. and MacPhee, A. G. and Magelssen, G. R. and Magoon, J. and Malone, R. M. and Malsbury, T. and Managan, R. and Mancini, R. and Manes, K. and Maney, D. and Manha, D. and Mannion, O. M. and Manuel, A. M. and Mapoles, E. and Mara, G. and Marcotte, T. and Marin, E. and Marinak, M. M. and Mariscal, C. and Mariscal, D. A. and Mariscal, E. F. and Marley, E. V. and Marozas, J. A. and Marquez, R. and Marshall, C. D. and Marshall, F. J. and Marshall, M. and Marshall, S. and Marticorena, J. and Martinez, D. and Maslennikov, I. and Mason, D. and Mason, R. J. and Masse, L. and Massey, W. and Masson-Laborde, P.-E. and Masters, N. D. and Mathisen, D. and Mathison, E. and Matone, J. and Matthews, M. J. and Mattoon, C. and Mattsson, T. R. and Matzen, K. and Mauche, C. W. and Mauldin, M. and McAbee, T. and McBurney, M. and Mccarville, T. and McCrory, R. L. and McEvoy, A. M. and McGuffey, C. and Mcinnis, M. and McKenty, P. and McKinley, M. S. and McLeod, J. B. and McPherson, A. and Mcquillan, B. and Meamber, M. and Meaney, K. D. and Meezan, N. B. and Meissner, R. and Mehlhorn, T. A. and Mehta, N. C. and Menapace, J. and Merrill, F. E. and Merritt, B. T. and Merritt, E. C. and Meyerhofer, D. D. and Mezyk, S. and Mich, R. J. and Michel, P. A. and Milam, D. and Miller, C. and Miller, D. and Miller, D. S. and Miller, E. and Miller, E. K. and Miller, J. and Miller, M. and Miller, P. E. and Miller, T. and Miller, W. and Miller-Kamm, V. and Millot, M. and Milovich, J. L. and Minner, P. and Miquel, J.-L. and Mitchell, S. and Molvig, K. and Montesanti, R. C. and Montgomery, D. S. and Monticelli, M. and Montoya, A. and Moody, J. D. and Moore, A. S. and Moore, E. and Moran, M. and Moreno, J. C. and Moreno, K. and Morgan, B. E. and Morrow, T. and Morton, J. W. and Moses, E. and Moy, K. and Muir, R. and Murillo, M. S. and Murray, J. E. and Murray, J. R. and Munro, D. H. and Murphy, T. J. and Munteanu, F. M. and Nafziger, J. and Nagayama, T. and Nagel, S. R. and Nast, R. and Negres, R. A. and Nelson, A. and Nelson, D. and Nelson, J. and Nelson, S. and Nemethy, S. and Neumayer, P. and Newman, K. and Newton, M. and Nguyen, H. and Di Nicola, J.-M. G. and Di Nicola, P. and Niemann, C. and Nikroo, A. and Nilson, P. M. and Nobile, A. and Noorai, V. and Nora, R. and Norton, M. and Nostrand, M. and Note, V. and Novell, S. and Nowak, P. F. and Nunez, A. and Nyholm, R. A. and O'Brien, M. and Oceguera, A. and Oertel, J. A. and Okui, J. and Olejniczak, B. and Oliveira, J. and Olsen, P. and Olson, B. and Olson, K. and Olson, R. E. and Opachich, Y. P. and Orsi, N. and Orth, C. D. and Owen, M. and Padalino, S. and Padilla, E. and Paguio, R. and Paguio, S. and Paisner, J. and Pajoom, S. and Pak, A. and Palaniyappan, S. and Palma, K. and Pannell, T. and Papp, F. and Paras, D. and Parham, T. and Park, H.-S. and Pasternak, A. and Patankar, S. and Patel, M. V. and Patel, P. K. and Patterson, R. and Patterson, S. and Paul, B. and Paul, M. and Pauli, E. and Pearce, O. T. and Pearcy, J. and Pedrotti, B. and Peer, A. and Pelz, L. J. and Penetrante, B. and Penner, J. and Perez, A. and Perkins, L. J. and Pernice, E. and Perry, T. S. and Person, S. and Petersen, D. and Petersen, T. and Peterson, D. L. and Peterson, E. B. and Peterson, J. E. and Peterson, J. L. and Peterson, K. and Peterson, R. R. and Petrasso, R. D. and Philippe, F. and Phipps, T. J. and Piceno, E. and Ping, Y. and Pickworth, L. and Pino, J. and Plummer, R. and Pollack, G. D. and Pollaine, S. M. and Pollock, B. B. and Ponce, D. and Ponce, J. and Pontelandolfo, J. and Porter, J. L. and Post, J. and Poujade, O. and Powell, C. and Powell, H. and Power, G. and Pozulp, M. and Prantil, M. and Prasad, M. and Pratuch, S. and Price, S. and Primdahl, K. and Prisbrey, S. and Procassini, R. and Pruyne, A. and Pudliner, B. and Qiu, S. R. and Quan, K. and Quinn, M. and Quintenz, J. and Radha, P. B. and Rainer, F. and Ralph, J. E. and Raman, K. S. and Raman, R. and Rambo, P. and Rana, S. and Randewich, A. and Rardin, D. and Ratledge, M. and Ravelo, N. and Ravizza, F. and Rayce, M. and Raymond, A. and Raymond, B. and Reed, B. and Reed, C. and Regan, S. and Reichelt, B. and Reis, V. and Reisdorf, S. and Rekow, V. and Remington, B. A. and Rendon, A. and Requieron, W. and Rever, M. and Reynolds, H. and Reynolds, J. and Rhodes, J. and Rhodes, M. and Richardson, M. C. and Rice, B. and Rice, N. G. and Rieben, R. and Rigatti, A. and Riggs, S. and Rinderknecht, H. G. and Ring, K. and Riordan, B. and Riquier, R. and Rivers, C. and Roberts, D. and Roberts, V. and Robertson, G. and Robey, H. F. and Robles, J. and Rocha, P. and Rochau, G. and Rodriguez, J. and Rodriguez, S. and Rosen, M. and Rosenberg, M. and Ross, G. and Ross, J. S. and Ross, P. and Rouse, J. and Rovang, D. and Rubenchik, A. M. and Rubery, M. S. and Ruiz, C. L. and Rushford, M. and Russ, B. and Rygg, J. R. and Ryujin, B. S. and Sacks, R. A. and Sacks, R. F. and Saito, K. and Salmon, T. and Salmonson, J. D. and Sanchez, J. and Samuelson, S. and Sanchez, M. and Sangster, C. and Saroyan, A. and Sater, J. and Satsangi, A. and Sauers, S. and Saunders, R. and Sauppe, J. P. and Sawicki, R. and Sayre, D. and Scanlan, M. and Schaffers, K. and Schappert, G. T. and Schiaffino, S. and Schlossberg, D. J. and Schmidt, D. W. and Schmitt, M. J. and Schneider, D. H. G. and Schneider, M. B. and Schneider, R. and Schoff, M. and Schollmeier, M. and Sch\"olmerich, M. and Schroeder, C. R. and Schrauth, S. E. and Scott, H. A. and Scott, I. and Scott, J. M. and Scott, R. H. H. and Scullard, C. R. and Sedillo, T. and Seguin, F. H. and Seka, W. and Senecal, J. and Sepke, S. M. and Seppala, L. and Sequoia, K. and Severyn, J. and Sevier, J. M. and Sewell, N. and Seznec, S. and Shah, R. C. and Shamlian, J. and Shaughnessy, D. and Shaw, M. and Shaw, R. and Shearer, C. and Shelton, R. and Shen, N. and Sherlock, M. W. and Shestakov, A. I. and Shi, E. L. and Shin, S. J. and Shingleton, N. and Shmayda, W. and Shor, M. and Shoup, M. and Shuldberg, C. and Siegel, L. and Silva, F. J. and Simakov, A. N. and Sims, B. T. and Sinars, D. and Singh, P. and Sio, H. and Skulina, K. and Skupsky, S. and Slutz, S. and Sluyter, M. and Smalyuk, V. A. and Smauley, D. and Smeltser, R. M. and Smith, C. and Smith, I. and Smith, J. and Smith, L. and Smith, R. and Sohn, R. and Sommer, S. and Sorce, C. and Sorem, M. and Soures, J. M. and Spaeth, M. L. and Spears, B. K. and Speas, S. and Speck, D. and Speck, R. and Spears, J. and Spinka, T. and Springer, P. T. and Stadermann, M. and Stahl, B. and Stahoviak, J. and Stanton, L. G. and Steele, R. and Steele, W. and Steinman, D. and Stemke, R. and Stephens, R. and Sterbenz, S. and Sterne, P. and Stevens, D. and Stevers, J. and Still, C. B. and Stoeckl, C. and Stoeffl, W. and Stolken, J. S. and Stolz, C. and Storm, E. and Stone, G. and Stoupin, S. and Stout, E. and Stowers, I. and Strauser, R. and Streckart, H. and Streit, J. and Strozzi, D. J. and Suratwala, T. and Sutcliffe, G. and Suter, L. J. and Sutton, S. B. and Svidzinski, V. and Swadling, G. and Sweet, W. and Szoke, A. and Tabak, M. and Takagi, M. and Tambazidis, A. and Tang, V. and Taranowski, M. and Taylor, L. A. and Telford, S. and Theobald, W. and Thi, M. and Thomas, A. and Thomas, C. A. and Thomas, I. and Thomas, R. and Thompson, I. J. and Thongstisubskul, A. and Thorsness, C. B. and Tietbohl, G. and Tipton, R. E. and Tobin, M. and Tomlin, N. and Tommasini, R. and Toreja, A. J. and Torres, J. and Town, R. P. J. and Townsend, S. and Trenholme, J. and Trivelpiece, A. and Trosseille, C. and Truax, H. and Trummer, D. and Trummer, S. and Truong, T. and Tubbs, D. and Tubman, E. R. and Tunnell, T. and Turnbull, D. and Turner, R. E. and Ulitsky, M. and Upadhye, R. and Vaher, J. L. and VanArsdall, P. and VanBlarcom, D. and Vandenboomgaerde, M. and VanQuinlan, R. and Van Wonterghem, B. M. and Varnum, W. S. and Velikovich, A. L. and Vella, A. and Verdon, C. P. and Vermillion, B. and Vernon, S. and Vesey, R. and Vickers, J. and Vignes, R. M. and Visosky, M. and Vocke, J. and Volegov, P. L. and Vonhof, S. and Von Rotz, R. and Vu, H. X. and Vu, M. and Wall, D. and Wall, J. and Wallace, R. and Wallin, B. and Walmer, D. and Walsh, C. A. and Walters, C. F. and Waltz, C. and Wan, A. and Wang, A. and Wang, Y. and Wark, J. S. and Warner, B. E. and Watson, J. and Watt, R. G. and Watts, P. and Weaver, J. and Weaver, R. P. and Weaver, S. and Weber, C. R. and Weber, P. and Weber, S. V. and Wegner, P. and Welday, B. and Welser-Sherrill, L. and Weiss, K. and Widmann, K. and Wheeler, G. F. and Whistler, W. and White, R. K. and Whitley, H. D. and Whitman, P. and Wickett, M. E. and Widmayer, C. and Wiedwald, J. and Wilcox, R. and Wilcox, S. and Wild, C. and Wilde, B. H. and Wilde, C. H. and Wilhelmsen, K. and Wilke, M. D. and Wilkens, H. and Wilkins, P. and Wilks, S. C. and Williams, E. A. and Williams, G. J. and Williams, W. and Williams, W. H. and Wilson, D. C. and Wilson, B. and Wilson, E. and Wilson, R. and Winters, S. and Wisoff, J. and Wittman, M. and Wolfe, J. and Wong, A. and Wong, K. W. and Wong, L. and Wong, N. and Wood, R. and Woodhouse, D. and Woodruff, J. and Woods, D. T. and Woods, S. and Woodworth, B. N. and Wooten, E. and Wootton, A. and Work, K. and Workman, J. B. and Wright, J. and Wu, M. and Wuest, C. and Wysocki, F. J. and Xu, H. and Yamaguchi, M. and Yang, B. and Yang, S. T. and Yatabe, J. and Yeamans, C. B. and Yee, B. C. and Yi, S. A. and Yin, L. and Young, B. and Young, C. S. and Young, C. V. and Young, P. and Youngblood, K. and Zacharias, R. and Zagaris, G. and Zaitseva, N. and Zaka, F. and Ze, F. and Zeiger, B. and Zika, M. and Zimmerman, G. B. and Zobrist, T. and Zuegel, J. D. and Zylstra, A. B.},
  collaboration = {Indirect Drive ICF Collaboration},
  journal = {Phys. Rev. Lett.},
  volume = {129},
  pages = {075001},
  year = {2022},
  doi = {10.1103/PhysRevLett.129.075001},
}

@Article{Turnbull_2020,
author={Turnbull, David
and Cola{\"i}tis, Arnaud
and Hansen, Aaron M.
and Milder, Avram L.
and Palastro, John P.
and Katz, Joseph
and Dorrer, Christophe
and Kruschwitz, Brian E.
and Strozzi, David J.
and Froula, Dustin H.},
title={Impact of the Langdon effect on crossed-beam energy transfer},
journal={Nature Physics},
year={2020},
month={Feb},
day={01},
volume={16},
number={2},
pages={181-185},
issn={1745-2481},
doi={10.1038/s41567-019-0725-z},
url={https://doi.org/10.1038/s41567-019-0725-z}
}

@article{Boehly_1999,
    author = {Boehly, T. R. and Smalyuk, V. A. and Meyerhofer, D. D. and Knauer, J. P. and Bradley, D. K. and Craxton,  R. S. and Guardalben, M. J. and Skupsky, S. and Kessler, T. J.},
    title = {Reduction of laser imprinting using polarization smoothing on a solid-state fusion laser},
    journal = {Journal of Applied Physics},
    volume = {85},
    number = {7},
    pages = {3444-3447},
    year = {1999},
    month = {04},
    issn = {0021-8979},
    doi = {10.1063/1.369702},
    url = {https://doi.org/10.1063/1.369702},
    eprint = {https://pubs.aip.org/aip/jap/article-pdf/85/7/3444/18979550/3444\_1\_online.pdf},
}

@article{Debayle_2025,
    author = {Debayle, A. and Loiseau, P. and Lecherbourg, L. and Masson-Laborde, P.-E. and Ruyer, C. and Morice, O. and Alozy, E. and Le-Deroff, L. and Caillaud, T. and Debesset, S. and Hermerel, C. and Rousseaux, C.},
    title = {Time-resolved cross-beam energy transfer in strongly damped regime on the Laser Mégajoule facility},
    journal = {Physics of Plasmas},
    volume = {32},
    number = {1},
    pages = {012705},
    year = {2025},
    month = {01},
    issn = {1070-664X},
    doi = {10.1063/5.0228994},
    url = {https://doi.org/10.1063/5.0228994},
    eprint = {https://pubs.aip.org/aip/pop/article-pdf/doi/10.1063/5.0228994/20358512/012705\_1\_5.0228994.pdf},
}

@article{lalaire,
author = {Lalaire, Y. and Ruyer, C. and Debayle, A. and Bouchard, G. and Fusaro, A. and Loiseau, P. and Masse, L. and Masson-Laborde, P.~E. and B\'enisti, D.},
title = {On the influence of optical smoothing techniques on cross-beam energy transfer},
    journal = {in. prep}, 
    volume = { },
    pages = { },
    year = {2025},
}

@article{Fried_1960,
doi = {10.1088/0368-3281/1/4/302},
url = {https://dx.doi.org/10.1088/0368-3281/1/4/302},
year = {1960},
month = {jan},
publisher = {},
volume = {1},
number = {4},
pages = {190},
author = {B D Fried and  M Gell-Mann and  J D Jackson and  H W Wyld},
title = {Longitudinal plasma oscillations in an electric field},
journal = {Journal of Nuclear Energy. Part C, Plasma Physics, Accelerators, Thermonuclear Research}
}

@article{Loiseau_2006,
author={P. Loiseau and O. Morice and D. Teychenn\'e and M. Casanova and S. H\"uller and D. Pesme},
title={Laser-beam smoothing induced by stimulated {Brillouin} scattering in an inhomogeneous plasma},
journal={Phys. Rev. Lett.},
volume={97},
pages={205001},
year={2006},
doi={10.1103/PhysRevLett.97.202001},
}

@article{Kritcher2022,
author = {Kritcher, Andrea and Young, Christopher and Robey, Harry and Weber, C. and Zylstra, A. and Hurricane, O. and Callahan, D. and Ralph, Joseph and Ross, J. and Baker, K. and Casey, D. and Clark, D. and Döppner, T. and Divol, L. and Hohenberger, Matthias and Hopkins, L. and Pape, S. and Meezan, N. and Pak, A. and Zimmerman, G.},
year = {2022},
month = {03},
pages = {},
title = {Design of inertial fusion implosions reaching the burning plasma regime},
volume = {18},
journal = {Nature Physics},
doi = {10.1038/s41567-021-01485-9}
}

@article{Oudin2022,
author = {Oudin,A.  and Debayle,A.  and Ruyer,C.  and Benisti,D. },
title = {Cross-beam energy transfer between spatially smoothed laser beams},
journal = {Physics of Plasmas},
volume = {29},
number = {11},
pages = {112112},
year = {2022},
doi = {10.1063/5.0109511},
URL = { https://doi.org/10.1063/5.0109511},
eprint = {https://doi.org/10.1063/5.0109511}
}

@article{Oudin2025,
author = {Oudin,A.  and Lalaire, Y. and Bouchard, G and Debayle,A.  and Fusaro, A. and Loiseau, P. and Ruyer,C.  and Benisti,D. },
title = {Theory and simulations of Cross-Beam Energy Transfer between speckled laser beams},
journal = {Physics of Plasmas},
volume = {29},
number = {00},
pages = {000000},
year = {2025},
}

@article{Seaton2022-1,
author = {Seaton,A. G.  and Yin,L.  and Follett,R. K.  and Albright,B. J.  and Le,A. },
title = {Cross-beam energy transfer in direct-drive ICF. I. Nonlinear and kinetic effects},
journal = {Physics of Plasmas},
volume = {29},
number = {4},
pages = {042706},
year = {2022},
doi = {10.1063/5.0078800},
URL = { 
        https://doi.org/10.1063/5.0078800 
},
eprint = { 
        https://doi.org/10.1063/5.0078800   
}
}

@article{Seaton2022-2,
author = {Seaton,A. G.  and Yin,L.  and Follett,R. K.  and Albright,B. J.  and Le,A. },
title = {Cross-beam energy transfer in direct-drive ICF. I. Nonlinear and kinetic effects},
journal = {Physics of Plasmas},
volume = {29},
number = {4},
pages = {042706},
year = {2022},
doi = {10.1063/5.0078800},

URL = { 
        https://doi.org/10.1063/5.0078800
    
},
eprint = { 
        https://doi.org/10.1063/5.0078800
    
}

}

@article{Neuville2018,
	doi = {10.1088/1361-6587/aaab23},
	url = {https://doi.org/10.1088%2F1361-6587%2Faaab23},
	year = 2018,
	month = {feb},
	publisher = {{IOP} Publishing},
	volume = {60},
	number = {4},
	pages = {044006},
	author = {C Neuville and K Glize and P Loiseau and P-E Masson-Laborde and A Debayle and M Casanova and C Baccou and C Labaune and S Depierreux},
	title = {Inhibition of crossed-beam energy transfer induced by expansion-velocity fluctuations},
	journal = {Plasma Physics and Controlled Fusion}
}

@article{Lefebvre_2018,
doi = {10.1088/1741-4326/aacc9c},
year = 2018,
volume = {59},
pages = {032010},
author = {E. Lefebvre and S. Bernard and C. Esnault and P. Gauthier and A. Grisollet and P. Hoch and L. Jacquet and G. Kluth and S. Laffite and S. Liberatore and I. Marmajou and P.-E. Masson-Laborde and O. Morice and J.-L. Willien},
title = {Development and validation of the {TROLL} radiation-hydrodynamics code for {3D} hohlraum calculations},
journal = {Nucl. Fusion},
}

@article{Garnier_1997,
doi = {10.1364/JOSAA.14.001928},
year = {1997},
volume = {14},
pages = {1928},
author = {J. Garnier and L. Videau and C. Gou\'edard and A. Migus},
title = {Statistical analysis for beam smoothing and some
applications},
journal = {J. Opt. Soc. Am. A},
}
\end{document}